\documentclass[12pt]{iopart}
\usepackage{graphicx, amssymb,amsthm, epsfig, amsbsy, amsfonts, setstack}
\usepackage[usenames,dvipsnames]{color}
\newtheorem{theorem}{Theorem}

\theoremstyle{definition}

\newcommand\I{\mathbf{I}}

\newcommand{\ui}{\mathrm{i}}
\newcommand{\ue}{\mathrm{e}}
\newcommand{\uD}{{Dir}}
\newcommand{\UI}{\mathbf{I}}

\newcommand{\ud}{\mathrm{d}}

\newcommand{\bbZ}{{\mathbb Z}}
\newcommand{\rz}{{\mathbb R}}
\newcommand{\bbN}{{\mathbb N}}

\newcommand{\re}{\Re}

\newcommand{\vp}{\boldsymbol{\psi}}

\newcommand{\vz}{\mathbf{0}}

\newcommand{\rk}{\mathrm{rank}}

\newcommand{\A}{{\mathbb A}}
\newcommand{\B}{{\mathbb B}}
\newcommand{\D}{{\mathbb D}}

\newcommand{\LM}{{\mathbf{L}}}
\newcommand{\bb}{{\overline{b}}}
\newcommand{\Sc}{{\mathcal{S}}}

\newcommand{\bdm}{\begin{displaymath}}
\newcommand{\edm}{\end{displaymath}}
\newcommand{\beq}{\begin{equation}}
\newcommand{\eeq}{\end{equation}}
\newcommand{\beqa}{\begin{eqnarray}}
\newcommand{\eeqa}{\end{eqnarray}}
\newcommand{\nn}{\nonumber}

\newcommand{\cL}{\mathcal{L}}

\def\I{{\rm i}}                  % le i mathematique
\def\D{{\rm d}}                  % la differenciation
\newcommand{\deriv}[2]{\frac{\mathrm{d}#1}{\mathrm{d}#2}}

\begin{document}

\title[Spectral determinants on graphs]{Spectral determinants and zeta functions of Schr\"odinger operators on metric graphs}

\author{JM Harrison$^1$, K Kirsten$^1$ and C Texier$^{2,3}$}

\address{$^1$ Department of Mathematics, Baylor University, Waco, TX 76798, USA}
\address{$^2$ Univ. Paris Sud; CNRS; LPTMS, UMR 8626, B\^at. 100, F-91405 Orsay, France}
\address{$^3$ Univ. Paris Sud; CNRS; LPS, UMR 8502, B\^at. 510, F-91405 Orsay, France}

\eads{\mailto{jon\_harrison@baylor.edu}, \mailto{klaus\_kirsten@baylor.edu}, \mailto{christophe.texier@u-psud.fr}}
\begin{abstract}
%Notes for a proof of the conjecture (61) of \cite{p:T:ZRSDMG} for the spectral determinant of a graph Schr\"odinger operator with general local vertex matching conditions.  We employ the technique introduced for graphs \cite{p:HK:ZQG}, first deriving the zeta function and from this the spectral determinant.
A derivation of the spectral determinant of the Schr\"odinger operator on a metric graph is presented where the local matching conditions at the vertices are of the general form classified according to the scheme of Kostrykin and Schrader.
To formulate the spectral determinant we first derive the spectral zeta function of the Schr\"odinger operator using an appropriate secular equation.
The result obtained for the spectral determinant is along the lines of the recent conjecture \cite{p:T:ZRSDMG}.
\end{abstract}

\submitto{\JPA}
\maketitle

\section{Introduction}
Quantum graphs have become a widely employed model for a range of phenomena in mathematical physics from Anderson localization and quantum chaos to problems in mesoscopic physics, microelectronics and the theory of waveguides, see e.g. \cite{p:K:GMWPTS, p:K:QG:IBS}.  Much of the success for the analysis of these problems on graphs can be attributed to the existence of well behaved spectral functions.  Consequently spectral determinants and trace formulae in graphs have become important problems widely studied in the literature \cite{p:ACDMT:SDQG, p:BK:TPSCSG, p:BE:TFQGGSABC, p:C:EPDG, ComDesTex05, p:D:SDGGBC, p:D:SDSOG, p:ES:BKOQG, p:ES:SIQG, p:F:DSOMG, p:GS:QG:AQCUSS, p:HK:ZQG, p:KMW:VDESDQSG, p:KPS:HKMGTF, p:KS:POTSSQG, p:R:SLSG, p:TM:QOMR}.
Here we consider the Sch\"odinger operator $-\triangle +V(x)$ on a metric graph where each bond of the $B$ bonds is associated with an interval $[0,L_b]$, $b=1,\dots, B$.
Matching conditions at the vertices of the graph are parameterized by $2B\times 2B$ matrices $\A$ and $\B$ according to the scheme introduced by Kostrykin and Schrader \cite{p:KS:KRQW}, see Section \ref{sec:graph model}.  We also restrict the possible matching conditions to \emph{local matching conditions} which both respect the graph topology, only relating values of the functions and their derivatives at the ends of intervals where they meet at a vertex, and are independent of the metric structure of the graph, namely the bond lengths $\{ L_b \}_{b=1,\dots,B}$ and the energy.
The matrices $\A$ and $\B$ can thus be considered to define the graph topology as they record how the bonds are connected.
%Local matching conditions are, hence, also independent of the energy.

In particular we will be concerned with the recent broad conjecture, introduced by one of the authors in \cite{p:T:ZRSDMG}, for the \emph{spectral determinant} which is defined formally as
\begin{equation}\label{eq:defn spec det}
    S(\gamma)=\det [\gamma -\triangle +V(x)]
    = {\prod_{j=0}^{\infty}} (\gamma+E_j) \ ,
\end{equation}
%
%\marginpar{\red{is ``$0<$''\\ neccessary ?}}
where $0< E_0 \leqslant E_1 \leqslant \dots $ is the spectrum of the Schr\"odinger operator and
$\gamma$ is some spectral parameter; the condition on the positivity of the spectrum can be relaxed by accepting additional technicalities \cite{p:KM:FDGSLP}.
The first result for the spectral determinant of the Schr\"odinger operator on a graph with general vertex matching conditions was obtained by Desbois \cite{p:D:SDGGBC}; the method was, however, unable to determine the $\gamma$-independent prefactor of the determinant. %In other terms the precise nature of the regularization of the determinant remained unclear in this first work.

In order to regularize the formal definition of $S(\gamma)$ we employ the \emph{spectral zeta function},
\begin{equation}\label{eq:spec zeta}
    \zeta(s)={\sum_{j=0}^{\infty}} (\gamma + E_j)^{-s} \ .
\end{equation}
The zeta function of the Schr\"odinger operator can be formulated by extending a technique introduced for graph Laplacians by some of the authors \cite{p:HK:VE, p:HK:ZQG}.
Having derived the zeta function the regularized spectral determinant is then defined by $S(\gamma)=\exp(-\zeta'(0))$.
More precisely we prove the following theorem along the lines of the conjecture.
\begin{theorem}\label{thm:spec det}
For the Schr\"odinger operator on a graph with local vertex matching conditions defined by a pair of matrices $\A$ and $\B$, with $\A\B^\dagger=\B\A^\dagger$ and $\rk (\A,\B)=2B$, and potential functions $V_b(x_b)\in C^\infty$ for $b=1,\dots, B$ the spectral determinant is
\end{theorem}
%Problem with \ui  =>  I put equation out of environment "theorem". Ch.
\bdm
    S(\gamma)= \left( \prod_{b=1}^B \frac{-2}{f'_{b} (L_b;\gamma)} \right)\frac{\det \left( \A +\B M(\gamma) \right)}{c_N \, \gamma^P} \ .
\edm
$f'_{b}(x;\gamma)$ and $M({\gamma})$ are defined in terms of the solution to a boundary value problem on the interval $[0,L_b]$ described in Section \ref{sec:secular}.  $c_N$ is the coefficient of the leading order $t\to \infty$ asymptotic behavior of $\det \left( \A +\B M(t^2) \right)$, determined in Section \ref{sec:asymp f(it)}.  Furthermore the exponent $P$ characterizes the $t\to0$ behavior of the determinant; $P=0$ if $\det \left( \A +\B M(0) \right)\neq 0$, the generic case, otherwise
$\det \left( \A +\B M(\gamma) \right) \sim \gamma^P$ as $\gamma \to 0$.

To complete the structure of the article, in Section  \ref{sec:graph model} we define the terminology of the quantum graph model.  Section \ref{sec:Dirichlet} introduces the technique, followed subsequently, in the simple case of the Schr\"odinger operator on an interval with Dirichlet boundary conditions and a graph with Dirichlet conditions at the ends of every bond.  To derive the zeta function of a general graph we start from a secular equation whose solutions are $\sqrt{E_j}$; an appropriate secular equation for the Schr\"odinger operator is identified in Section \ref{sec:secular}.  In Section \ref{sec:zeta} we formulate the graph zeta function with general local matching conditions at the vertices.  Finally, in Section \ref{sec:det}, we apply the results for the zeta function to prove Theorem \ref{thm:spec det}.  The Appendix shows the derivation of the $t\to \infty$ asymptotics of the function $f'_{b}(x;t^2)$.

\section{Graph model}\label{sec:graph model}
A graph $G$ is a collection of \emph{vertices} $v=1,\dots,V$ and \emph{bonds} $b=1,\dots,B$.
%see Figure \ref{fig:graph}.
Each bond connects a pair of vertices $b=(v,w)$ and we denote by $o(b)=v$ the initial vertex of $b$ and $t(b)=w$ the terminal vertex.  $\bb = (w,v)$ will denote the reversed bond with the initial and terminal vertices exchanged.  We consider undirected graphs with $B$ undirected bonds, so $b$ and $\bb$ refer to the same physical bond but the availability of two labels for each bond will be notationally convenient.  To be explicit the label $b$ will refer to the bond with $o(b)<t(b)$.
$m_v$ will denote the number of bonds meeting at a vertex $v$, the valency of $v$.
To determine the valency $b$ and $\overline{b}$ refer to a single bond.

%\begin{figure}[!ht]
%  \begin{center}
%  \includegraphics[width=4cm]{graphexample}
%  \caption{\it A typical finite graph.}
%  \label{fig:graph}
%  \end{center}
%\end{figure}

In order to define a metric graph we associate with each bond $b$ an interval $[0,L_b]$ where $L_b$ is the length of $b$ and $x_b=0$ at $o(b)$ and $x_b=L_b$ at $t(b)$.
So $x_b$ is the distance to a point in the interval measured from the initial vertex $o(b)$ and we also use $x_{\bb}=L_b-x_b$ for the distance to the same point measured from the terminal vertex of $b$.
The total length of $G$ is denoted by $\cL=\sum_{b=1}^B L_b$.
A function $\psi$ on $G$ is defined by specifying the set of functions $\{ \psi_b(x_b) \}_{b=1,\dots,B}$ on the collection of intervals.
The redundancy of notation enforces the relation $\psi_b(x_b)=\psi_\bb(x_\bb)=\psi_\bb(L_b-x_b)$.
The Hilbert space of $G$ is consequently
\begin{equation}\label{eq:Hilbert space}
    {\mathcal H} = \bigoplus_{b=1}^B L^2 \bigl( [0,L_b] \bigr) \ .
\end{equation}

Motivated by physical applications we consider here Schr\"odinger operators on the metric graph.   The eigenproblem on bond $b$ is
\begin{equation}\label{eq:eigenproblem}
    \left( \ui \frac{\ud}{\ud x_b} +A_b \right)^2 \psi_b(x_b) + V_b(x_b) \psi_b(x_b) = k^2 \psi_b(x_b) \ .
\end{equation}
The set $\{A_1,\dots,A_B\}$ defines a vector potential on the graph constant on each bond.\footnote[1]{A vector potential $A_b(x_b)$ that depends on the position on the bond can be made constant on each bond by a gauge transformation.} For consistency $A_\bb=-A_b$ as the direction is reversed when changing coordinate.
%The scalar potential $V_b$ is assumed to be symmetric, hence $V_b(x_b)=V_b(L_b-x_b)$.

\section{Dirichlet determinants}\label{sec:Dirichlet}
\subsection{Laplace operator on a finite interval}
We first analyse the determinant of the Laplace operator on a single finite
interval $[0,L]$ with Dirichlet boundary conditions, a \emph{wire}.
The spectrum of the wire is $E_j=\big(\frac{j\pi}{L}\big)^2$ for
$j\in\mathbb{N}^*$ and we can compute the
zeta-function directly,
\begin{equation}
  \zeta_\mathrm{Dir}(s) = \sum_{j=1}^\infty E_j^{-s}
  = \left(\frac{L}{\pi}\right)^{2s} \zeta_R(2s) \ .
\end{equation}
Differentiating,
\begin{equation}
    \zeta_\mathrm{Dir}'(0) = 2\zeta_R'(0) + 2\zeta_R(0)\,\log(L/\pi)
  = -\log2 - \log L \ ,
\end{equation}
where we have used $\zeta_R(0)=-\frac12$ and $\zeta_R'(0)=-\frac12\log(2\pi)$
\cite{gragra}.
We therefore obtain the spectral determinant for $\gamma=0$,
\begin{equation}\label{eq:Dir det wire V=0}
  S_\mathrm{Dir}(0) = \det(-\Delta) = \exp \Big(- \zeta_\mathrm{Dir}'(0) \Big) =
  2L \ ,
\end{equation}
which agrees with the formulation of the spectral determinant of the
Laplace operator in \cite{p:HK:ZQG, p:KL:CDCI, p:T:ZRSDMG}.

\subsection{Schr\"odinger operator on a finite interval}
To compute the spectral determinant of the operator
$-\deriv{^2}{x^2}+V(x)$ for $x\in[0,L]$ with Dirichlet boundary conditions
let us introduce two useful basis for the solutions of the differential
equation
\begin{equation}
  \label{eq:SE}
  \left(-\deriv{^2}{x^2}+V(x)\right)\psi(x) = k^2 \, \psi(x)
  \:.
\end{equation}
The first basis
shown to be useful in~\cite{p:D:SDSOG, p:D:SDGGBC} is:
%is guided by the conjecture of \cite{p:T:ZRSDMG}:
$f(x;-k^2)$ is a solution of (\ref{eq:SE}) such that $f(0;-k^2)=1$ and $f(L;-k^2)=0$,
with the second linearly independent solution denoted
$\bar f(\bar x;-k^2)$ where $\bar x=L-x$ and $\bar f$ satisfies
$\bar f(0;-k^2)=1$ and $\bar f(L;-k^2)=0$.
For a symmetric potential, $V(x)=V(L-x)$ and we have $\bar f(x;-k^2)=f(x;-k^2)$.
An alternative basis was employed in \cite{p:FKM:PMP}:
$u_{k}(x;-k^2)$ and $\bar u(\bar x;-k^2)$ are solutions satisfying
$u(0;-k^2)=0$ and $u'(0;-k^2)=1$,
where prime denotes derivation with respect to the spatial coordinate $x$, i.e. first argument of the functions. The relation between the two basis is
given by
\begin{equation}
    \bar u(\bar x;-k^2) = { - \frac{ f(x;-k^2) }{ f'(L;-k^2) } }
   \hspace{1cm}\mbox{or}\hspace{1cm}
   f(x;-k^2) = \frac{ \bar u(\bar x;-k^2) }{ \bar u(L;-k^2) }
   \:.
\end{equation}
A secular equation whose solutions $k$ are square roots of the eigenvalues of the Schr\"odinger operator  is then
%[Ref]\footnote{{\sf It should be a standard result in this simple case...What Ref?}}
\begin{equation}
  \label{eq:SecularEquation}
   u(L;-k^2) = 0
  \: ;
\end{equation}
note that
the secular equation may be written in different forms, thanks to
the relations
$u(L;-k^2) = \bar u(L;-k^2)
= -1/ f'(L;-k^2) = -1/ \bar f'(L;-k^2)$,
that hold for an arbitrary potential. Note, however, that $u(x;\gamma)$
and $\bar u(x;\gamma)$ do not coincide in general (apart for a
symmetric potential $V(x)=V(L-x)$), nevertheless
$u(L;\gamma)=\bar u(L;\gamma)$ always holds.  This may be shown by
computing the Wronskian of the two solutions at the
ends of the interval.

\paragraph{Example 1:}
For a free particle where $V(x)=0$,
\begin{equation}\label{eq:f Dir V=0}
\fl
   f(x;-k^2) = \bar f(x;-k^2) = \frac{\sin k(L-x) }{\sin kL}
   \hspace{0.5cm}\mbox{and}\hspace{0.5cm}
   u(x;-k^2) = \bar u(x;-k^2) = \frac1{k}\sin kx \ .
\end{equation}
The Dirichlet spectrum is then given by solutions of the secular equation
\begin{equation}\label{eq:wire secular}
   {F(k)=}  u(L;-k^2)=\frac{-1}{f'(L;-k^2)}=\frac1k\sin kL=0 \ .
\end{equation}

\paragraph{Example 2:}
In the case of a linear potential $V(x)=\omega\,x$
we find
\begin{eqnarray}
\label{eq:FctUAiry}
\fl  u(x;-k^2)
= \frac{\pi}{\omega^{1/3}}
  \left[
    \mathrm{Ai}(-\omega^{-2/3}k^2)\,
    \mathrm{Bi}(\omega^{1/3}(x-k^2/\omega)) \right. \nn \\
    \hspace*{4.5cm} \left.
   -\mathrm{Bi}(-\omega^{-2/3}k^2)\,
    \mathrm{Ai}(\omega^{1/3}(x-k^2/\omega))
   \right] \ ,\\
\fl  \bar u(\bar x;-k^2) = \frac{\pi}{\omega^{1/3}}
  \left[
    \mathrm{Bi}(\omega^{1/3}(L-k^2/\omega))\,
    \mathrm{Ai}(\omega^{1/3}(x-k^2/\omega)) \right. \nn \\
   \hspace*{4.5cm} \left. -\mathrm{Ai}(\omega^{1/3}(L-k^2/\omega))\,
    \mathrm{Bi}(\omega^{1/3}(x-k^2/\omega))
   \right] \ ,
\end{eqnarray}
with the Airy functions $\mathrm{Ai}(z)$ and $\mathrm{Bi}(z)$ \cite{b:AS:HMF}.
%The spectrum is then obtained by solving $u_{k}(L)=0$.
The Dirichlet spectrum is given by solving the secular equation
\begin{equation}
 \fl
 F(k)=\mathrm{Ai}(-\omega^{-2/3}k^2)\mathrm{Bi}(\omega^{1/3}(L-k^2/\omega))
  - \mathrm{Bi}(-\omega^{-2/3}k^2)\mathrm{Ai}(\omega^{1/3}(L-k^2/\omega))
%  \frac{ \mathrm{Ai}(-\omega^{-2/3}k^2) }{ \mathrm{Bi}(-\omega^{-2/3}k^2) } -
%  \frac{ \mathrm{Ai}(\omega^{1/3}(L-k^2/\omega)) }{ \mathrm{Bi}(\omega^{1/3}(L-k^2/\omega)) }
  = 0
   \ .
\end{equation}

\paragraph{Example 3:}
Another interesting example is the case of a potential of the form
$V(x)=\phi(x)^2+\phi'(x)$. The Hamiltonian
\begin{equation}
  H_S=-\deriv{^2}{x^2}+\phi(x)^2+\phi'(x)
\end{equation}
describes so-called supersymmetric quantum mechanics \cite{Wit81}.
In this case, exploiting the factorization $H_S=Q^\dagger Q$ with $Q=-\deriv{}{x}+\phi(x)$, it is possible to construct explicitly two linearly independent solutions of the differential equation $H_S\psi=0$. We may choose
$\psi_0(x)=\exp\int_0^x\phi(y) \, \D y$
and
$\psi_1(x)=\psi_0(x)\int_x^L\frac{\D y}{\psi_0(y)^2}$.
We obtain the two useful solutions for $k=0$:
\begin{equation}
  \label{eq:FunctionsUsusy}
  \fl
  u(x;0) = \psi_0(0)\psi_0(x)\int^x_0\frac{\D y}{\psi_0(y)^2}
   \hspace{0.5cm}\mbox{and}\hspace{0.5cm}
  \bar u(\bar x;0) = \psi_0(L)\psi_0(x)\int^L_x\frac{\D y}{\psi_0(y)^2}
\end{equation}

\vspace{0.25cm}

\begin{figure}[!ht]
  \begin{center}
  \setlength{\unitlength}{1cm}
    \begin{picture}(14,4.5)
    \put(0,0){\includegraphics[scale=0.9]{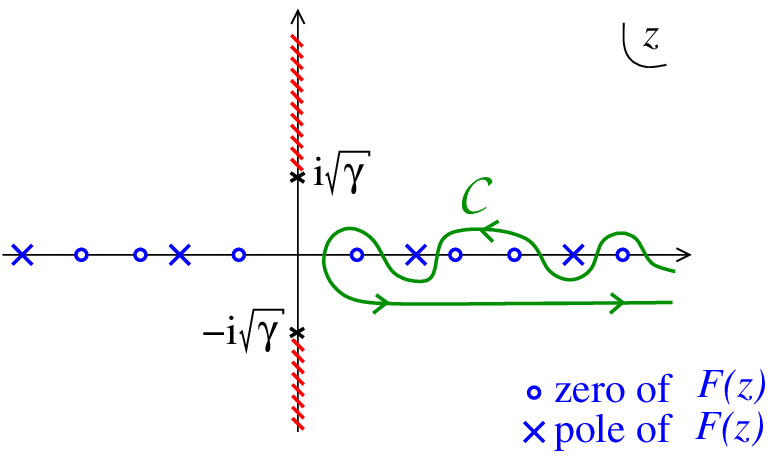}}
    \put(7,0){\includegraphics[scale=0.9]{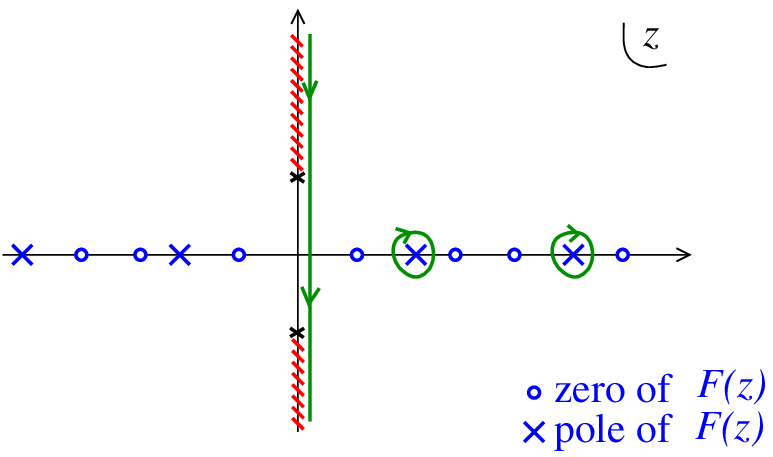}}
    \put(0,4){a)}
    \put(7,4){b)}
    \end{picture}
  \caption{\it The contours used to evaluate the graph zeta function,
    (a) before, and (b) after, the contour deformation.
    The two branch cuts are represented by hashed lines.}
  \label{fig:contours}
  \end{center}
\end{figure}

We now explain the general method followed in \cite{p:HK:ZQG}
and in the present article.
A secular equation $F({k})=0$ can be related to
the zeta-function using the argument principle,
\begin{equation}
  \zeta_\mathrm{Dir}(s,\gamma)
  = \frac{1}{2\I\pi} \int_\mathcal{C}
  (z^2+\gamma)^{-s}\deriv{}{z}\log\left( F(z) \right) \, \D z \ ,
\end{equation}
where the contour wraps around the positive real axis enclosing the
roots of $F(z)$, the solutions of the secular equation, and avoiding
any poles, Figure \ref{fig:contours}(a).

In the present case the function $F(z)=u(L;-z^2)$ does not present
poles. However, as there is some freedom in the choice of $F(z)$, in the following {section} it is convenient to choose a function $F$ for the
general graph that will possess poles.
Transforming the contour $\mathcal{C}$ to the imaginary axis $z=\ui t$, Figure \ref{fig:contours}(b), we obtain
\begin{equation}
   \zeta_\mathrm{Dir} (s,\gamma)  = \frac{\sin\pi s}{\pi}
  \int_{\sqrt{\gamma}}^\infty
  (t^2-\gamma)^{-s}\,
  \deriv{}{t}\log\left( u(L;t^2) \right)\, \D t
  \:,
\end{equation}
where, assuming $F(0)\neq 0$, the segment between $\I\sqrt{\gamma}$ and $-\I\sqrt{\gamma}$ does
not contribute to the integral as $F(z)=F(-z)$ (which
follows from the symmetry of the {secular equation}); comments about the case $F(0)=0$ follow in Section \ref{sec:det}.
%spectrum).
Analysis of the integral shows that this representation of the
zeta-function is valid in the strip $1/2<\re s<1$.
The restriction to $\re s >1/2$ comes from the asymptotic behavior of $u(x;t^2)$ for $t\to \infty$.  The asymptotics of the function $u(x;t^2)$ for $t\to \infty$ were evaluated in~\cite{p:FKM:PMP},
\begin{equation}
  \log   u(L;t^2) \underset{t\to\infty}{\sim}
  t L - \log 2t + O(t^0)
  \:.
\end{equation}
An explicit example of such an expansion is provided in the appendix, see Eq.~(\ref{eq:fasymp2}).
To obtain a representation valid for $s\to 0$ we
subtract and add the first two terms in the asymptotic
expansion.   Then upon integration we find,
\begin{eqnarray}
  \fl
  \zeta_\mathrm{Dir}(s,\gamma)  = \frac{\sin\pi s}{\pi}
  \int_{\sqrt{\gamma}}^\infty
  (t^2-\gamma)^{-s}\,
  \deriv{}{t}
  \left[
     \log\left( u(L;t^2) \right) - t L + \log(2t)
  \right]\, \D t \nn \\
  \hspace*{4cm} + L \frac{\Gamma(s-1/2)}{2\sqrt\pi \Gamma(s)}\gamma^{\frac12-s}
  -\frac12\gamma^{-s}
  \:,
\end{eqnarray}
valid in the strip $-1/2<\re s<1$.
Differentiating
\begin{equation}
\fl  \zeta'_\mathrm{Dir}(0,\gamma)
   = -\log\left( u(L;\gamma) \right)
  - \log(2\sqrt\gamma) + \frac12\log\gamma
   = -\log\left(2 u(L;\gamma) \right)
  \:.
\end{equation}
Consequently the zeta-regularized spectral determinant of a wire with Dirichlet
boundary conditions is \cite{p:KM:FDPZM},
\begin{equation}
  S_\mathrm{Dir}(\gamma) = \exp\Big( -\zeta'_\mathrm{Dir}(0,\gamma) \Big)
  = 2 u(L;\gamma) = \frac{-2}{f'(L;\gamma)}
  \:.
\end{equation}

\paragraph{Example 1:}
In the absence of a potential we obtain
\begin{equation}
  \label{eq:NoV}
  S_\mathrm{Dir}(\gamma) = \frac{2\sinh\sqrt\gamma L}{\sqrt\gamma}
  \:.
\end{equation}
As $\gamma\to 0$ this expression agrees with (\ref{eq:Dir det wire V=0}).

\paragraph{Example 2:}
Using (\ref{eq:FctUAiry}), the spectral determinant of $H=-\deriv{^2}{x^2}+\omega x$ therefore is
\begin{equation}
\fl
  S_\mathrm{Dir}(\gamma) =
  \frac{2\pi}{\omega^{1/3}}
  \left[
    \mathrm{Ai}(\omega^{-2/3}\gamma)\,
    \mathrm{Bi}(\omega^{1/3}(L+\gamma/\omega))
   -\mathrm{Bi}(\omega^{-2/3}\gamma)\,
    \mathrm{Ai}(\omega^{1/3}(L+\gamma/\omega))
   \right]
  \:.
\end{equation}

\paragraph{Example 3:}
Using the expression (\ref{eq:FunctionsUsusy}) we can obtain an explicit formula for the spectral determinant of the supersymmetric Hamiltonian for vanishing spectral parameter:
\begin{equation}
  S_\mathrm{Dir}(0) = 2\,\psi_0(0)\psi_0(L)\int^L_0\frac{\D y}{\psi_0(y)^2} \ ,
\end{equation}
i.e.
\begin{equation}
   \label{eq:DirDetSusy}
  \det\left( -\deriv{^2}{x^2}+\phi^2+\phi' \right)
  = 2\, e^{\int_0^L\phi(x)\, \D x}\int_0^Le^{-2\int_0^x\phi(y)\, \D y}\, \D x
  \:.
\end{equation}
Note that this explicit formula furnishes another way to obtain the determinant
$\det(\gamma-\deriv{^2}{x^2}+V)$ in terms of one solution among a family of solutions of a Riccati equation
$\phi'(x)=\gamma-\phi(x)^2+V(x)$.
For example, in the free case, $V(x)=0$, the Riccati problem is solved by
$\phi(x)=\sqrt\gamma\tanh\sqrt\gamma(x+x_0)$: we can check that (\ref{eq:DirDetSusy}) leads to (\ref{eq:NoV}).
Formula (\ref{eq:DirDetSusy}) may also be illustrated in the case
$\phi(x)=\omega x$ (harmonic oscillator on a finite interval),
\begin{equation}
  \label{eq:DetHarmOscFiniteInterval}
   \det\left( -\deriv{^2}{x^2}+\omega^2 x^2+\omega \right)
  = \sqrt{\frac{\pi}{\omega}} \, e^{\frac12 \omega L^2}\,
  \mathrm{erf}(\sqrt\omega L)
  \:,
\end{equation}
where $\mathrm{erf}(x)$ is the error function \cite{gragra}.
The limit $L\to\infty$ of
(\ref{eq:DetHarmOscFiniteInterval}) is singular and does not approach either
$\sqrt{\pi/\omega}$ or $\sqrt{\pi/2\omega}$, the determinant of the
operator defined on $\mathbb{R}$ or $\mathbb{R}_+$ respectively.
This is not unexpected as eigenfunctions on the interval $[0,L]$ will not be square integrable in the limit $L\to \infty$.  Eigenfunctions of high enough energy always feel the presence of the boundary
conditions at both ends of the interval and thus differ significantly from the infinite space situation.

\subsection{Dirichlet determinant of a graph}
We are now also in a position to consider the case of a graph with Dirichlet boundary conditions
at every vertex. %the ends of every bond.
The Dirichlet conditions describe a graph where all the bonds are decoupled and consequently
the spectrum is simply the union of the Dirichlet spectra of the
individual intervals.
Since the zeta-functions of each wire are additive and
the determinants
are multiplicative in this case we recover a result of \cite{p:T:ZRSDMG}
\begin{equation}\label{eq:Dir spec det}
  S_\mathrm{Dir}(\gamma) =
  \prod_{b=1}^B \frac{-2}{f'_{b}(L_b;\gamma)}
  \:.
\end{equation}

To compare with the free case,
we use (\ref{eq:f Dir V=0}) to get
$S_\mathrm{Dir}(\gamma) =
2^B\gamma^{-B/2}\prod_{b=1}^B  \sinh\sqrt{\gamma} L_b$.
%we recall (\ref{eq:f Dir V=0})
%$-f'_{b}(L_b;\gamma)^{-1}=\gamma^{-1/2} \sinh \sqrt{\gamma} L_b$.
Taking the limit $\gamma \to 0$ we see (\ref{eq:Dir spec det}) reduces
to the explicit formula $S_\mathrm{Dir}(0) = 2^B\prod_{b=1}^B L_b$
which agrees with (\ref{eq:Dir det wire V=0}).

\section{Secular equation of a Schr\"odinger operator with general matching conditions}\label{sec:secular}
For a general graph let $f_{b}$ be the solution of the boundary value
problem on the interval $[0,L_b]$,
\begin{equation}\label{eq:interval boundary problem}
    \left( -\frac{\ud^2}{\ud x_b^2} + V_b(x_b) \right) f_{b} (x_b;-k^2) = k^2 f_{b} (x_b;-k^2) \ ,
\end{equation}
such that
$f_{b}(0;-k^2)=1$ and $f_{b}(L_b;-k^2)=0$.
 $f_{\bb}(x_\bb;-k^2)$
will denote the complimentary solution on the same bond where the function vanishes at the initial vertex of $b$ and is unity at the terminal vertex, i.e. $f_{\bb}(0;-k^2)=1$ and $f_{\bb}(L_b;-k^2)=0$.\footnote[2]{Notice that it is convenient to consider $f_{\bb}$ as a function of $x_{\bb}=L_b-x_b$.}  $f_b$ and $f_\bb$ are a pair of linearly independent functions on the interval satisfying the eigenvalue problem.   Hence the component of the wavefunction of the Schr\"odinger operator (\ref{eq:eigenproblem}) with energy $k^2$ on bond $b$ can then be written
\begin{equation}\label{eq:wavefn on b}
    \psi_{b} (x_b,-k^2) = c_b \, f_{b} (x_b;-k^2) \, \ue^{\ui A_b x_b}  + c_\bb \, f_{\bb} (x_\bb;-k^2)
    \, \ue^{\ui A_\bb  x_\bb} \ .
\end{equation}

Matching conditions at the graph vertices are specified by a pair of $2B\times 2B$ matrices $\A$ and $\B$ via
\begin{equation}\label{eq:matching conditions}
    \A \vp + \B \hat{\vp}=\vz \ ,
\end{equation}
where
\begin{eqnarray}\label{eq:vertex values}
    \vp&=&\big(\psi_1(0),\dots,\psi_B(0),\psi_1(L_1),\dots,\psi_B(L_B)\big)^T \ , \\
    \hat{\vp}&=&\big(D_{1} \psi_1(0),\dots, D_{B} \psi_B(0), D_{\overline{1}} \psi_1(L_1),\dots,D_{\overline{B}} \psi_B(L_B)\big)^T \ .
\end{eqnarray}
$D_b := \frac{\ud}{\ud x_b} -\ui A_b$ is the covariant derivative and in (\ref{eq:vertex values}) we consider $\psi_b$ a function of $x_b$ so $x_b=0$ at $o(b)$ and $x_b=L_b$ at $t(b)$.  $\hat{\vp}$ is therefore the vector of inward pointing covariant derivatives at the ends of the intervals.

The following theorem of Kostrykin and Schrader which classifies all matching conditions of self-adjoint realizations of the Laplace operator \cite{p:KS:KRQW} also applies to the Schr\"odinger operator.
\begin{theorem}\label{thm:sa matching conditions}
The Laplace operator with matching conditions specified by $\A$ and $\B$ is self-adjoint if and only if $(\A,\B)$ has maximal rank and $\A\B^\dagger=\B\A^\dagger$.
\end{theorem}
See \cite{p:K:QG:I} for an alternative unique classification scheme for general vertex matching conditions.  Note that the general matching conditions we employ can be approximated by ornamenting the vertices of the graph with subgraphs whose vertices have $\delta$-type matching conditions \cite{p:CET:AGSVC}.

We consider only \emph{local matching conditions} where the matrices $\A$ and $\B$ relate values of functions and their derivatives on the intervals where they meet at a vertex and where the matrices are independent of the metric structure of the graph, namely the bond lengths and the spectral parameter $k$.  The restriction on the form of matching conditions is necessary in order to employ the argument principle later.

Representing the components of an energy eigenfunction on $G$ following (\ref{eq:wavefn on b}) we see that
$\vp=(c_1,\dots,c_B,c_{\overline{1}},\dots, c_{\overline{B}} )^T$ and $\hat{\vp}=M(-k^2)\vp$ where $M(-k^2)$ is the $2B\times 2B$ matrix
\begin{equation}\label{eq:defn of M}
    M_{ab}=\delta_{a,b} \, f'_{b}(0;-k^2) - \delta_{a,\bb} \, f'_{\bb } (L_b;-k^2) \ue^{\ui A_b L_b} \ .
\end{equation}
The vector $\vp$ of coefficients of the two linearly independent solutions on each bond characterizes the wavefunction.  Substituting in (\ref{eq:matching conditions}) we see that for $\vp$ to define a wavefunction of energy $k^2$ satisfying the matching conditions we require
\begin{equation}\label{eq:secular}
    \det \left( \A +\B M(-k^2) \right) =0 \ ,
\end{equation}
which defines a secular equation for the graph Schr\"odinger operator.

\paragraph{Example 1:}
For a free particle $V_b(x_b)=0$ and we can write $f_{b}$ explicitly,
\begin{equation}\label{eq: V=0 g}
\fl    f_{b}(x_b;-k^2)= \frac{\sin k(L_b-x_b)}{\sin k L_b} \ , \qquad \textrm{ so } \qquad
    f_{b}'(x_b;-k^2)= \frac{-k\cos k(L_b-x_b)}{\sin k L_b} \ .
\end{equation}
%
%and
%
%\begin{equation}\label{eq: V=0 g'}
%  g_{b,k}'(L_b)= \frac{-k\cos k(L_b-x_b)}{\sin k L_b} \ .
%\end{equation}
%
Consequently the matrix $M$ takes the form
\begin{equation}\label{V=0 M}
    M(-k^2)= -k \left( \begin{array}{cc}
    \cot k \LM & -\csc k \LM \\
    - \csc k \LM & \cot k \LM \\
    \end{array} \right) \ ,
\end{equation}
where $\cot k \LM =\textrm{diag} \{ \cot k L_1 , \dots, \cot k L_B \}$ and $\csc k \LM$ is defined similarly.
%
%It will be useful for future reference, to note that, for a graph without potential,
%
%\begin{equation}\label{eq: V=0 g'(L)}
%    f_{b,\ui \sqrt{\gamma}}'(L_b)^{-1}= \frac{\sinh \sqrt{\gamma} L_b}{\sqrt \gamma} \ .
%\end{equation}

\section{Zeta function of a Schr\"odinger operator with general matching conditions}\label{sec:zeta}
Let us define the function
\begin{equation}\label{eq:f}
    F(z):=\det \left( \A +\B M(-z^2) \right)
\end{equation}
so the secular equation (\ref{eq:secular}) reads $F(k)=0$.  As previously the zeta function is written as a contour integral
\begin{equation}
    \zeta(s,\gamma) = \frac{1}{2\pi \ui} \int_{\mathcal{C}} (z^2+\gamma)^{-s} \frac{\ud}{\ud z} \log F(z) \,  \ud z \ ,
    \label{eq:zeta contour int}
\end{equation}
where the contour $\mathcal{C}$ encloses the zeros of $F$ and avoids its poles, Figure \ref{fig:contours}(a).
Making a contour transformation to the imaginary axis, Figure \ref{fig:contours}(b), the integral breaks into two parts
\begin{equation}\label{eq:Im plus poles}
    \zeta(s,\gamma)=\zeta_\mathrm{Im}(s,\gamma) +\zeta_{\mathrm{P}}(s,\gamma) \ ,
\end{equation}
 where we denote by $\zeta_\mathrm{Im}$ the integral along the imaginary axis and $\zeta_\mathrm{P}$ is the series that results from subtracting the residue $(z_0^2+\gamma)^{-s}$ at each pole $z_0$ of $F$.
Writing $z=\ui t$ the imaginary axis integral becomes
\begin{equation}\label{eq:zeta fn s>1/2}
    \zeta_\mathrm{Im}(s,\gamma)%&=& \frac{1}{2\pi \ui} \int_\infty^{-\infty} (\gamma-t^2)^{-s} \frac{\ud}{\ud t} \log f(\ui t) \,  \ud t \nn \\
    = \frac{\sin \pi s}{\pi} \int_{\sqrt{\gamma}}^{\infty} (t^2-\gamma)^{-s} \frac{\ud}{\ud t} \log F(\ui t) \,  \ud t \ , \label{eq:zeta Im}
\end{equation}
which converges in the strip $1/2<\re s<1$.

As poles of $F$ lie at eigenvalues of the graph Schr\"odinger operator with Dirichlet boundary conditions we see that $\zeta_{\mathrm{P}}(s,\gamma)=\zeta_\mathrm{Dir}(s,\gamma)$, which we evaluated previously.  It remains to formulate the imaginary axis integral so that the zeta function given in (\ref{eq:zeta fn s>1/2}) is well defined to the left of the line $\re s = 1/2$.  %To achieve this we must understand the $t\to \infty$ asymptotics of $F(\ui t)$.  %As a first step we therefore analyze the asymptotics of $g'_{\ui t}$, we have dropped the subscript $b$ for the time being to make the notation more compact.

\subsection{Large $t$ asymptotics of $F(\ui t)$}\label{sec:asymp f(it)}
To extend the zeta function to the left of the line $\re s = 1/2$ we require the large $t$ asymptotics of $F(\ui t)=\det (\A+\B M(t^2))$.
The $t\to \infty$ asymptotics of $f'_{b}$ are analysed in the appendix where we find
\begin{equation}\label{eq:g'(L) asymp early}
    f'_{b}(0;t^2) \sim  -t +\sum_{j=1}^\infty s_{b,j}(0)t^{-j} \ ,
\end{equation}
and $f'_b(L_b;t^2)$ vanishes exponentially.  The functions $s_{b,j}(x_b)$ are determined iteratively from the potential $V_b(x_b)$ via a recursion relation (\ref{eq:recursion}). %They can be determined to an arbitrary order using a computer algebra package.

Using (\ref{eq:g'(L) asymp early}) we can write the expansion of $M(t^2)$ as $t\to \infty$ up to exponentially suppressed terms.
\begin{equation}\label{eq:M expansion}
     M_{ab}\sim \delta_{a,b}(-t+\sum_{j=1}^\infty s_{b,j}(0)t^{-j}) \ .
\end{equation}
Hence
\begin{equation}\label{eq:f expansion}
    F(\ui t) \sim \det \Big(\A+\B(-t\UI+D(t))\Big)
\end{equation}
where
\begin{equation}\label{eq:D(t)}
  D(t)= \sum_{j=1}^\infty t^{-j} \textrm{diag} \left\{   s_{1,j}(0),\dots, s_{B,j}(0), s_{\overline{1},j}(0),\dots, s_{\overline{B},j}(0) \right\} \ .
\end{equation}
The expansion of $F(\ui t)$ fits with the free particle case investigated in \cite{p:HK:ZQG} where, in the absence of a potential, $D(t)=\vz$ and $F(\ui t) \sim \det (\A-t\B)$.

Using (\ref{eq:f expansion}) and the recursion relation (\ref{eq:recursion}) the expansion of $F(\ui t )$ can be carried out to arbitrary order using a computer algebra package.  In general the expansion will take the form
\begin{equation}\label{eq:generic F expansion}
    F(\ui t) \sim \det \Big(\A+\B(-t\UI+D(t))\Big) = \sum_{j=0}^\infty c_j \,  t^{2B-j} \ ,
\end{equation}
where $c_0=\det \B$.  We denote by $c_N$ the first nonzero coefficient in the expansion of {$F(\ui t)$}, so $N=0$ if $\det \B \neq 0$.  To evaluate the zeta-regularized spectral determinant we need a representation of the zeta function valid about $s=0$.  To find this we only require the leading order in the expansion.  Subtracting and adding the leading order behavior we obtain an expression for the imaginary axis integral valid in the strip $-1/2 < \re s < 1$.
\begin{equation}
\fl %    \zeta_\mathrm{Im}(s, \gamma) = \frac{\sin \pi s}{\pi} \int_{\sqrt{\gamma}}^{\infty} t^{-2s} \frac{\ud}{\ud t} \log \big( F(\ui t)t^{N-2B} /c_N \big) \,  \ud t + \frac{(2B-N) \sin \pi s}{2\pi s} \gamma^{-s} \ .
%\blue{
  \zeta_\mathrm{Im}(s, \gamma) = \frac{\sin \pi s}{\pi} \int_{\sqrt{\gamma}}^{\infty} (t^2-\gamma)^{-s} \frac{\ud}{\ud t} \log \big( F(\ui t)t^{N-2B} /c_N \big) \,  \ud t + \frac12(2B-N) \gamma^{-s} \ .
%}
\label{eq:zeta Im extended}
\end{equation}
Correspondingly
\begin{eqnarray}\label{eq:zeta' Im}
    \zeta_\mathrm{Im}'(0, \gamma)&=\log c_N -\log \big( F(\ui \sqrt{\gamma}) \gamma^{\frac{N-2B}{2}} \big) -\frac{2B-N}{2} \log \gamma \, \nn \\
    &= \log c_N - \log \det \left( \A +\B M({\gamma}) \right) \ .
\end{eqnarray}

\section{Zeta-regularized spectral determinant}\label{sec:det}
Collecting the results so far we can now formulate the spectral determinant, $S(\gamma) = \exp\big(-\zeta'_\mathrm{Dir}(0,\gamma)-\zeta'_\mathrm{Im}(0,\gamma)\big)$, of a graph with general local vertex matching conditions, we obtain,
\begin{equation}\label{eq:general spec det}
\fl  S(\gamma)= \frac{S_\mathrm{Dir}(\gamma)}{c_N} \det \left( \A +\B M(\gamma) \right) \qquad \textrm{where} \qquad S_\mathrm{Dir}(\gamma)=\prod_{b=1}^B \frac{-2}{f'_{b} (L_b;\gamma)} \ .
\end{equation}
This is the statement of Theorem \ref{thm:spec det}.
{The $\gamma$-dependent part of this expression coincides with Desbois' result \cite{p:D:SDGGBC} obtained by a different approach.}
If we choose Dirichlet matching conditions at the graph vertices then $\A=\UI_{2B}$ and $\B$ is a matrix of zeros, see (\ref{eq:matching conditions}).  Hence $c_{2B}=1$, which is the only non-zero term in the large $t$ expansion of {$F(\ui t)$}, and $\det \left( \A +\B M({\gamma}) \right)=1$, consequently (\ref{eq:general spec det}) reduces to (\ref{eq:Dir spec det}) as required.

\paragraph{$F(z)$ vanishing at zero.}  If $F(0)=0$ the contour integral will pick up contributions associated with the behavior at $z=0$.  While $F(0)$ does not vanish generically, when it does vanish it is straightforward to compensate for this by adjusting the function $F$.  Assume $F(z)\sim z^{2P}$ as $z\to 0$ for $P\in \mathbb{N}^*$.\footnote[1]{The even power is due to the symmetry of $F$.}  We may define a new function $\tilde{F}(z)=F(z)/z^{2P}$.  Then $\tilde{F}(z)=0$ is a secular equation but $\tilde{F}$ does not vanish at zero.  Evaluating the zeta function using $\tilde{F}$ the procedure is identical and we find,
\begin{equation}
  S(\gamma)= \frac{S_\mathrm{Dir}(\gamma)}{c_N\, \gamma^P} \det \left( \A +\B M(\gamma) \right) \ .
\end{equation}
The same technique was used, for example, to evaluate the zeta function of the Laplacian on a star graph with Neumann like matching conditions in \cite{p:HK:ZQG}.
To determine $P$ in a particular case one requires the asymptotic expansion of $f'_b(0,t^2)$ and $f'_b(L_b,t^2)$ as $t \to 0$.  %It is worth pointing out that
We now discuss the spectral determinant in two particular cases.

\paragraph{Functions continuous at the vertices.}
We consider local vertex matching conditions for which the
wavefunction is continuous at the vertices, namely the $\delta$-type
conditions.
At a vertex $v$ the local $\delta$-type matching conditions relate the
functions and covariant derivatives at the ends of the intervals
meeting at $v$. One explicit choice of $m_v\times m_v$ matrices to
encode such matching conditions is
\begin{equation}
\fl  \mathbb{A}_v  =
  \left(
    \begin{array}{ccccc}
      -\lambda_v & 0 & 0 & \cdots & 0\\
      -1 & 1 & 0 & \cdots & 0\\
      0 & -1 & 1 & \cdots & 0\\
      \vdots & \vdots &  & & \vdots \\
      0 & 0 & 0 & \cdots  & 1
    \end{array}
  \right)
  \hspace{0.5cm} \mbox{and} \hspace{0.5cm}
  \mathbb{B}_v  =
  \left(
    \begin{array}{ccccc}
     1 & 1 & \cdots & 1 & 1\\
     0 & 0 & \cdots & 0 & 0 \\
     0 & 0 & \cdots & \cdots & 0 \\
     \vdots & \vdots &  & & \vdots \\
     0 & 0 & \cdots & \cdots & 0
    \end{array}
  \right)
\end{equation}
where $\lambda_v$ is the strength of the coupling at $v$.  We note that
$\det(\mathbb{A}_v-t\mathbb{B}_v)=-(\lambda_v+m_vt)$.
By relabeling the $2B$ ends of the intervals we may write the matrices $\A$ and  $\B$ describing matching conditions on the whole graph so the matrices $\A_v$ and  $\B_v$ for $v=1,\dots,V$ appear as blocks on the diagonal.  We recall that the factor $c_N$ appearing in the spectral determinant is the coefficient of the highest power of $t$ in the expansion of $\det (\A +\B (-t\UI +D(t) )$, where $D(t)$ is a diagonal matrix where each non-zero element is a series in inverse powers of $t$.  Hence we see that for a graph with $\delta$-type conditions at all the vertices and any potentials on the bonds
\begin{equation}
\fl  \det(\mathbb{A}+\mathbb{B}M({\gamma}))
  \underset{t\to\infty}{\sim}
  \det (\A +\B (-t\UI +D(t) )
  \underset{t\to\infty}{\sim} \left(\prod_v -m_v\right) t^V +O(t^{V-1})
  \:.
\end{equation}
Thus we find $N=V$ and  $c_V=\prod_v (-m_v)$.  Consequently
\begin{equation}
  \label{eq:SDC}
  S(\gamma) = \left(\prod_{b=1}^B \frac{-2}{f'_{b}(L_b;\gamma)}\right)
  \frac{\det(\mathbb{A} + \mathbb{B} M({\gamma}))}{\prod_v (-m_v)}
  \:,
\end{equation}
reproducing the result in \cite{p:T:ZRSDMG}.
{Note that the choice of a wavefunction continuous at the vertices allows one to introduce vertex variables and simplify the spectral determinant by replacing
$\det(\mathbb{A} + \mathbb{B} M({\gamma}))$ by a matrix coupling vertices of smaller dimension \cite{p:ACDMT:SDQG,p:D:SDGGBC,p:T:ZRSDMG}.}

\paragraph{Functions whose derivative is continuous at the vertices.}
Another interesting case of matching conditions, studied in
\cite{Tex08,p:T:ZRSDMG}, are the so called $\delta_s'$-type matching
conditions corresponding to a wavefunction whose derivative is
continuous at the vertices \cite{Exn95}.
A possible choice of matching matrices at a vertex $v$ is
\begin{equation}
\fl  \mathbb{A}_v  =
  \left(
    \begin{array}{ccccc}
     1 & 1 & \cdots & 1 & 1\\
     0 & 0 & \cdots & 0 & 0 \\
     0 & 0 & \cdots & \cdots & 0 \\
     \vdots & \vdots &  & & \vdots \\
     0 & 0 & \cdots & \cdots & 0
    \end{array}
  \right)
  \hspace{0.5cm} \mbox{and} \hspace{0.5cm}
  \mathbb{B}_v  =
  \left(
    \begin{array}{ccccc}
      -\mu_v & 0 & 0 & \cdots & 0\\
      -1 & 1 & 0 & \cdots & 0\\
      0 & -1 & 1 & \cdots & 0\\
      \vdots & \vdots &  & & \vdots \\
      0 & 0 & 0 & \cdots  & 1
    \end{array}
  \right) \ .
\end{equation}
So
$\det(-\frac1t\mathbb{A}_v+\mathbb{B}_v)=-(\mu_v+\frac1tm_v)$,
therefore
\begin{equation}
  \det(\mathbb{A}+\mathbb{B}M({\gamma}))
  \underset{t\to\infty}{\sim}
  t^{2B-V}\prod_v (-m_v)
   \hspace{1cm}\mbox{for } \mu_v=0 \:\forall \, v
  \:.
\end{equation}
The exponent $N=2B-V$ and the coefficient
$c_{2B-V}=\prod_v (-m_v)$. We again obtain (\ref{eq:SDC}).

\paragraph{Comparison with the conjecture of \cite{p:T:ZRSDMG}.}
The zeta-regularized spectral determinant was obtained in
\cite{p:T:ZRSDMG} in the two particular cases considered above with a
different method.
Based on the observation that both sets of matching conditions led to
Eq.~(\ref{eq:SDC}) it was conjectured that this result is of greater
generality (Eq.~(61) of this reference).
In light of the present work we now inject two observations.
\begin{itemize}
\item The matrices $\mathbb{A}$ and $\mathbb{B}$ describing boundary
  conditions are not unique. Any transformation
  $(\mathbb{A},\,\mathbb{B})\to(U\mathbb{A},\,U\mathbb{B})$ where $U$
  is a matrix such that $\det U\neq0$,
%(this condition is sufficient  but maybe not necessary),
leaves the nature of the matching conditions
  invariant. Therefore the general expression of the zeta-regularized
  spectral determinant should be invariant under such a transformation.
  It is clear that this is not the case for the form (\ref{eq:SDC})
  conjectured in \cite{p:T:ZRSDMG} whereas our main result
  Theorem \ref{thm:spec det} is invariant.
\item The conjecture was based on a continuity argument with
  respect to variations of $\mathbb{A}$ and $\mathbb{B}$.
  However, in general, $c_N$ may depend on the particular choice of the
  matching conditions rather than simply the topology of the graph. In particular, it is not possible to rule out, a
    priori, that for a given choice of matching conditions and
  potentials a term in the series expansion (\ref{eq:generic F expansion}) vanishes.
  We have seen in the previous pair of examples that the exponent $N$
  does indeed vary with the nature of the matching conditions.
\end{itemize}

\ack
The authors would like to thank Alain Comtet, Sebastian Eggers n\'e Endres, Guglielmo Fucci, Frank Steiner and Yves Tourigny for stimulating discussions.
KK is supported by National Science Foundation grant PHY--0757791.  JMH was supported by the Baylor University summer sabbatical program.
%\blue{CT acknowledges stimulating discussions with Alain Comtet and Yves Tourigny.}

\appendix\section*{Appendix. Large $t$ asymptotics of $f'(x;t^2)$}
\setcounter{section}{1}
$f(x;t^2)$ is the solution of
\begin{equation}\label{eq:transformed eigenproblem}
    \left( -\frac{\ud^2}{\ud x^2} +V(x) +t^2 \right) f(x;t^2) = 0
\end{equation}
on the interval $[0,L]$ with $f(0;t^2) = 1$ and $f(L;t^2) = 0$.  To find the asymptotics of $f'(x;t^2)$ in the limit $t\to \infty$ we let $\phi_t(x)$ be a general solution of (\ref{eq:transformed eigenproblem}) and call $\Sc_t(x)=\frac{\ud}{\ud x} \log \phi_t(x)$. $\Sc_t$ satisfies the differential equation
\begin{equation}\label{eq:dif eq for S}
    \Sc_t'(x)=t^2+V(x)-\Sc_t^2(x) \ .
\end{equation}
Consequently, in the limit $t\to \infty$, $\Sc_t$ has an expansion of the form
\begin{equation}\label{eq:expansion of S}
    \Sc_t(x)=\sum_{j=-1}^\infty s_j(x) \, t^{-j} \ .
\end{equation}
The functions $s_j(x)$ are determined by the recurrence relation
\begin{equation}\label{eq:recursion}
  s_{j+1} (x)= \mp \frac{1}{2} \left( s_j'(x) +\sum_{k=0}^j s_k(x) s_{j-k}(x) \right)
\end{equation}
where $s_{-1}(x) = \pm 1$, $s_0(x) = 0$ and  $s_1(x) = \pm \frac{1}{2} V(x)$.
The two alternative signs in the expansion produce exponentially growing and decaying solutions of (\ref{eq:transformed eigenproblem}).  If we denote with $\Sc^{\pm}_t(x)$ the functions with $s_{-1}(x)=\pm 1$ respectively then we may write the large $t$ behavior of $f(x;t^2)$ as a linear combination of the exponentially growing and decaying parts,
\begin{equation}\label{eq:assymptotic g}
    f(x;t^2) =A^+ \exp \int_0^x \Sc^+_t(u) \, \ud u  + A^- \exp \int_0^x \Sc^-_t(u) \, \ud u  \ .
\end{equation}
Imposing the boundary conditions we require $A^-=1-A^+$ and as it turns out
\begin{equation}\label{eq:g asymp bcs}
    A^\pm = \frac{\mp \exp \int_0^L \Sc^-_t(u) \, \ud u}{\exp \int_0^L \Sc^\mp_t(u) \, \ud u - \exp \int_0^L \Sc^-_t(u) \, \ud u } \ .
\end{equation}
Consequently differentiating (\ref{eq:assymptotic g}) we find,
\begin{eqnarray}\label{eq:g' asymp}
\fl     f'(x;t^2)=  \\
     \fl \frac{\Sc^-_t(x) \exp  (\int_0^x \Sc^-_t(u) \, \ud u + \int_0^L \Sc^+_t(u) \, \ud u) - \Sc^+_t(x) \exp  (\int_0^x \Sc^+_t(u) \, \ud u  + \int_0^L \Sc^-_t(u) \, \ud u)}{\exp \int_0^L \Sc^+_t(u) \, \ud u - \exp \int_0^L \Sc^-_t(u) \, \ud u} \ . \nn
\end{eqnarray}

We are interested in the $t\to \infty$ asymptotics of both $f'(0;t^2)$ and $f'(L;t^2)$.  From (\ref{eq:g' asymp}) we see that at $x=L$, $f'(x;t^2)$ vanishes exponentially.  At $x=0$, up to exponentially suppressed terms, we have the asymptotic relation
\begin{equation}\label{eq:g'(L) asymp}
    f'(0;t^2) \sim \Sc^{-}_t(0) = -t +\sum_{j=1}^\infty s_j(0)t^{-j} \ .
\end{equation}

\paragraph{Example:}
In the case of the linear potential $V(x)=\omega\,x$ considered earlier
the first eight functions $s_j$ in the expansion of $\Sc^-_t$ constructed iteratively from the recurrence relation are shown bellow.
\begin{eqnarray*}
  s_{-1}(x)=-1 \qquad  \qquad s_0(x)=0 \qquad & s_1(x)=-\frac{\omega}{2}x \nn \\
  s_2(x)=-\frac{\omega}{2^2} \qquad \qquad s_3(x)=\frac{\omega^2}{2^3}x^2 \qquad & s_4(x)=\frac{\omega^2}{2^2}x \nn \\
  s_5(x) =\frac{3\omega^2}{2^5}-\frac{\omega^3}{2^3}x^3 \qquad & s_6(x) =-\frac{11\omega^3}{2^5} x^2  \nn
\end{eqnarray*}
Consequently
\begin{equation}\label{eq:fasymp2}
    f'(0;t^2) \sim  -t -\frac{\omega}{2^2 t^2} +\frac{3\omega^2}{2^5t^5}
    + \cdots \ .
\end{equation}

\end{document}